\documentclass[12pt]{article}

\usepackage{amsmath}
\usepackage{amssymb}
\usepackage{amsfonts}
\usepackage{latexsym}
\usepackage{color}
\usepackage{graphicx}

\catcode `\@=11 \@addtoreset{equation}{section}

\catcode `\@=12

\newcommand{\be}{\begin{equation}}
\newcommand{\en}{\end{equation}}
\newcommand{\bea}{\begin{eqnarray}}
\newcommand{\ena}{\end{eqnarray}}
\newcommand{\beano}{\begin{eqnarray*}}
\newcommand{\enano}{\end{eqnarray*}}
\newcommand{\bee}{\begin{enumerate}}
\newcommand{\ene}{\end{enumerate}}

\newcommand{\mc}{\mathcal}

\newcommand{\D}{{\mc D}}

\newcommand{\Sc}{{\cal S}}
\newcommand{\E}{{\cal E}}
\newcommand{\F}{{\cal F}}
\newcommand{\G}{{\cal G}}
\newcommand{\K}{{\mathfrak K}}

\newcommand{\Lc}{{\cal L}}

\newcommand{\1}{1 \!\! 1}

\newcommand{\Hil}{\mc H}
\newcommand{\kt}{\rangle}
\newcommand{\br}{\langle}

\newtheorem{thm}{Theorem}

\newtheorem{defn}[thm]{Definition}

\catcode `\@=11 \@addtoreset{equation}{section}
\catcode `\@=12

\begin{document}

\thispagestyle{empty}

\vspace*{2cm}

\begin{center}
{\Large \bf Coupled Susy, pseudo-bosons and a deformed $\mathfrak{su}(1,1)$ Lie algebra}   \vspace{2cm}\\

{\large F. Bagarello}\\
  Dipartimento di Ingegneria,
Universit\`a di Palermo,\\ I-90128  Palermo, Italy\\
and I.N.F.N., Sezione di Napoli\\
e-mail: fabio.bagarello@unipa.it\\
home page: www1.unipa.it/fabio.bagarello

\end{center}

\vspace*{2cm}

\begin{abstract}
\noindent In a recent paper a pair of operators $a$ and $b$ satisfying the equations $a^\dagger a=bb^\dagger+\gamma\1$ and  $aa^\dagger=b^\dagger b+\delta\1$, has been considered, and their nature of ladder operators has been deduced and analysed. Here, motivated by the spreading interest in non self-adjoint operators in Quantum Mechanics, we extend this situation to a set of four operators, $c$, $d$, $r$ and $s$, satisfying $	dc=rs+\gamma\1$ and $cd=sr+\delta\1$, and we show that they are also ladder operators. We show their connection with biorthogonal families of vectors and with the so-called $\D$-pseudo bosons. Some examples are discussed.
\end{abstract}

\vspace{2cm}


\vfill


\newpage

\section{Introduction}

Coupled SUSY (CSusy) was introduced in \cite{coupledsusy} in connection with quantum mechanical systems in presence of some particular ladder operators. A simple example of CSusy is provided by the quantum harmonic oscillator, but other examples are discussed in \cite{coupledsusy}. Many relevant systems described in terms of ladder operators have been discussed in the literature during many years. We refer here to the monograph \cite{dong}, and to the papers \cite{lad1}-\cite{lad5}, which are just few of those written in the past three years.
This very partial list of references is sufficient to give an idea of the vitality of the topic, also in contexts which might appear quite non standard, see for instance \cite{bagbook1} and \cite{bagbook2}.

It is well known that ladder operators are not uniquely related to canonical commutation, or anti-commutation, rules. Raising and lowering operators also appear for quons, \cite{fiv,green,kar}, and in the analysis of the truncated harmonic oscillator, \cite{buc,bagchi}, just to cite two other interesting situations. As already observed, they also appear in CSusy, \cite{coupledsusy}, where they are proved to be related to the $\mathfrak{su}(1,1)$ Lie algebra.

In almost all the cases listed above, these operators are used to write an Hamiltonian of the physical system, and to deduce its dynamics out of it. The Hamiltonian, and more in general the observables of the system, are taken to be self-adjoint. This is, as it is well known, the standard approach in Quantum Mechanics, \cite{mess,merz}. However, since some decades, it has become clearer and clearer that self-adjointness of an operator is a sufficient condition for its eigenvalues to be real. But it is not also necessary. And this has physical consequences. In fact, it is trivial to construct counterexamples: it is enough to consider a non self-adjoint operator $T$ which is similar (but not unitarily equivalent) to a self-adjoint
$T_0$. This means that an invertible operator $V$ exists such that $T=VT_0V^{-1}$. Here, to make the situation simple, we assume that all the operators are bounded, $T,T_0,V,V^{-1}\in B(\Hil)$, the C*-algebra of the bounded operators over the Hilbert space $\Hil$. In this case, if $\varphi$ is an eigenstate of $T_0$ with eigenvalue $E\in\mathbb{R}$, it is clear that $V\varphi$ is different from zero, and that it is an eigenstate of $T$ with the same eigenvalue. Furthermore, if $V^{-1}\neq V^\dagger$, then $T\neq T^\dagger$. This simple consideration, with a seminal paper on the now famous Hamiltonian $H=p^2+ix^3$, \cite{ben1}, turned on a renewed interest to physicists  for non self-adjoint observables\footnote{Mathematicians were already aware of the possibility of having real eigenvalues for non self-adjoint operators.}. A very rich line of research was then generated, involving people interested in the new physical consequences of this approach, and people more focused on its mathematical aspects, which are many and not so easy to deal with. For instance, even if the reality of the eigenvalues is preserved, the eigenvectors need not being orthonormal, or even being a basis in $\Hil$, also in presence of a pure point spectrum. Some references on the physical aspects of what is now usually called {\em PT} or {\em Pseudo-Hermitian} Quantum Mechanics are \cite{ben1}-\cite{bagprocpa}, while \cite{bagbookPT}-\cite{petr4} are references more mathematically oriented.

This partial list of references prove that there is a growing interest in a deeper understanding of the properties of many classes of non self-adjoint operators, and in particular of some connected {\em extended number-like operators} and their related ladder operators. This analysis has begun several years ago when people realized that, in some physically relevant situation, it is possible to factorize a certain Hamiltonian $H$, non self-adjoint, as follows: $H=A^*A$. Here $A^*$ is not the (Dirac) adjoint of $A$, $A^\dagger$, and this explains why $H\neq H^\dagger$. This property has been analyzed in details in different contexts along the years for several different systems, producing deformed versions of the standard {\em particles}: $\D$-pseudo bosons ($\D$-PBs), pseudo-fermions, deformed quons, and other deformations of {\em ordinary} bosons, fermions and quons. The common feature, in all these cases, is that the raising operator of the model is not the Dirac adjoint of the lowering operator. In principle, they are unrelated, even if they usually satisfy some peculiar (anti-)commutation rules. We refer to \cite{baginbagbook,bagthmp} for some self-consistent reviews on these particular families of deformations.

In this paper we continue our analysis, starting from CSusy and deforming the related ladder operators. In this way we consider two different (but related) families of four raising and four lowering operators, acting on different vectors which are biorthogonal in pairs and connected by these operators. They are also eigenvectors of four different number-like, manifestly non self-adjoint, operators. This is essentially the content of Section \ref{sect2} where, after introducing the problem, we construct the algebraic framework where all the operators considered all along the paper live. After creating the settings, we deform CSusy and we discuss some useful aspects of the deformed $\mathfrak{su}(1,1)$ Lie algebra arising in our analysis. The supersymmetric aspects of the model are discussed in Section \ref{sectBECSusy}. In Section \ref{sectPBs} we show how $\D$-PBs produce an interesting example of our construction, while in Section \ref{sectDPBs} we consider a deformed version of $\D$-PBs, with an explicit example connected to the shifted harmonic oscillator, \cite{petr4}. Our conclusions are given in Section \ref{sectconcl}.

\section{The settings}\label{sect2}

In \cite{coupledsusy} the authors introduced the notion of {\em coupled SUSY} (CSusy), and they used it in the analysis of some physical systems with Hamiltonians written in terms of ladder operators. Roughly speaking, a CSusy arises out of two operators $a$ and $b$, acting on an Hilbert space $\Hil$, and two real non-zero numbers $\gamma$, $\delta$, with $\delta>\gamma$, satisfying the following:
\be
a^\dagger a=bb^\dagger+\gamma\1, \qquad aa^\dagger=b^\dagger b+\delta\1.
\label{21}\en
 Here $\1$ is the identity operator on $\Hil$. Since $a$, $b$ and their adjoints could be unbounded, it is clear that some domain conditions must be imposed to these operators. For instance, for (\ref{21}) to make sense, the range of $a$ ($b$) must be contained in the domain of $a^\dagger$ ($b^\dagger$), and viceversa. It is easy to see that (\ref{21}) extends the ordinary bosonic case. In fact, if $c$ is an operator on $\Hil$ satisfying (in the sense on unbounded operators) the canonical commutation relation $[c,c^\dagger]=\1$, then the equations in (\ref{21}) are satisfied taking $a=b=c$, $\delta=1$, $\gamma=-1$.

  The interesting result deduced in \cite{coupledsusy} is that most of the typical ladder structure {\em attached} to $c$ and $c^\dagger$, producing in particular an orthonormal basis in $\Hil$, can be recovered also using the pair $(a,b)$ in (\ref{21}). In particular, the operators
 \be
 \K_+=\frac{1}{\delta-\gamma}a^\dagger b^\dagger, \qquad \K_-=\frac{1}{\delta-\gamma} ba, \qquad \K_0=\frac{1}{\delta-\gamma}\left(a^\dagger a-\frac{\gamma}{2}\right),
\label{22}\en
satisfy the following commutation rules, 
\be
[\K_0,\K_\pm]=\pm\K_\pm, \qquad [\K_+,\K_-]=-2\K_0,
\label{23}\en
which are those of the $\mathfrak{su}(1,1)$ Lie algebra. Hence, $\K_\pm$ act as ladder operators, while $\K_0$ is connected with some Hamiltonian.

Equations (\ref{22}) show that $\K_0=\K_0^\dagger$ and $\K_+^\dagger=\K_-$. Our interest is focused in what happens when these conditions are lost. In particular, we are interested in considering what happens if the ladder operators extending $\K_\pm$ are not one the adjoint of the other, and if $\K_0\neq \K_0^\dagger$. As already discussed in the Introduction, this become interesting in view of the growing interest for physical systems driven by manifestly non self-adjoint Hamiltonians, \cite{ben1}-\cite{bagbookPT}, which has produced several results both in theoretical and experimental physics, and in pure mathematics. To produce this extension, we will replace (\ref{21}) with two new equalities involving four, in general unrelated, operators $c$, $d$, $r$ and $s$, all acting on $\Hil$. However, before doing this, it is convenient to introduce an algebraic settings which can be useful to deal with our situation, creating a rather general framework.

\subsection{$O^*$-algebras}\label{sectalgebras}

Since the original proposal by R. Haag and H. Kastler in 1964, \cite{hk}, it become clear that the use of algebras of operators can be useful in the analysis of several physical systems, and in particular of those with infinite degrees of freedom. However, $C^*$ or von Neumann-algebras are sometimes not the best choice, since the operators they contain are all bounded, while many physical systems are deeply connected with operators which are not bounded. This is quite often the case in many-body theory, in quantum field theory, and in statistical mechanics, for instance, \cite{br1}-\cite{sewbook2}. But it is also true for simple quantum mechanical systems. For instance, the ladder operators used in the analysis of the quantum harmonic oscillator, and their related number operator, are all unbounded, \cite{br2,llt,araki}. To deal with these cases, in the past twenty years or so several examples of {\em unbounded operator algebras} have been introduced and studied in details. We refer to \cite{aitbook}-\cite{Inoue} for some relevant publications on this subject. 
In this paper we will use a particular unbounded operator algebra, the $O^*$-algebra  $\Lc^\dagger(\D)$.

Let us briefly review how  $\Lc^\dagger(\D)$ can be introduced, and let us comment on why it is so relevant for us. We start with the following definition:

\begin{defn}\label{o*}Let $\mathcal{H}$ be a separable Hilbert space and $N_0$ an
	unbounded, densely defined, self-adjoint operator. Let $D(N_0^k)$ be
	the domain of the operator $N_0^k$, $k \ge 0$, and $\mathcal{D}$ the domain of
	all the powers of $N_0$, that is,  \be \mathcal{D} =  \bigcap_{k\geq 0}
	D(N_0^k). \label{add1}\en This set is dense in $\mathcal{H}$. We call
	$\mathcal{L}^\dagger(\mathcal{D})$ the $*$-algebra of all \textit{  closable operators}
	defined on $\mathcal{D}$ which, together with their adjoints, map $\mathcal{D}$ into
	itself. Here the adjoint of $X\in\mathcal{L}^\dagger(\mathcal{D})$,
	$X^\dagger$, is the restriction of the adjoint of $X$ in $\Hil$ (which we also indicate with $X^\dagger$) to $\D$. 
\end{defn}

In $\mathcal{D}$ the topology is defined by the following $N_0$-depending
seminorms: $$\phi \in \mathcal{D} \rightarrow \|\phi\|_n\equiv \|N_0^n\phi\|,$$
where $n \ge 0$, while  the topology $\tau_0$ in $\mathcal{L}^\dagger(\mathcal{D})$ is introduced by the seminorms
$$ X\in \mathcal{L}^\dagger(\mathcal{D}) \rightarrow \|X\|^{f,k} \equiv
\max\left\{\|f(N_0)XN_0^k\|,\|N_0^kXf(N_0)\|\right\},$$ where
$k \ge 0$ and   $f \in \mathcal{C}$, the set of all the positive,
bounded and continuous functions  on $\mathbb{R}_+$, which are
decreasing faster than any inverse power of $x$:
$\mathcal{L}^\dagger(\mathcal{D})[\tau_0]$ is a {   complete *-algebra}. This implies, in particular, that taken any
 $x,y\in \mathcal{L}^\dagger(\mathcal{D})$, we can multiply them and the results, $xy$ and $yx$, both belong to $\mathcal{L}^\dagger(\mathcal{D})$, as well as their difference, the commutator $[x,y]$. Also, powers of $x$ and $y$ all belong to $\Lc^\dagger(\D)$, which  therefore is a good candidate to work with, also in presence of unbounded operators. In fact, if $N_0=c^\dagger c$, where $[c,c^\dagger]=\1$ as above, we can prove that $c, c^\dagger\in\Lc^\dagger(\D)$. Hence $N_0\in  \Lc^\dagger(\D)$ as well.

Let now $a$ and $b$ be two operators
on $\mathcal{H}$, with domains $D(a)$ and $D(b)$ respectively, $a^\dagger$ and $b^\dagger$ their adjoint, and let $\mathcal{D}$ be a dense subspace of $\mathcal{H}$
such that $a^\sharp\mathcal{D}\subseteq\mathcal{D}$ and $b^\sharp \mathcal{D} \subseteq \mathcal{D}$, where with $x^\sharp$ we indicate $x$ or $x^\dagger$. Of course, $\mathcal{D}\subseteq D(a^\sharp)$
and $\mathcal{D}\subseteq D(b^\sharp)$.

\begin{defn}\label{def21}
	The operators $(a,b)$ are $\mathcal{D}$-\textit{pseudo-bosonic}  if, for all $f\in\mathcal{D}$, we have
	\begin{equation}\label{A1}
	a\,b\,f-b\,a\,f=f.
	\end{equation}
\end{defn}
 
By means of $a$ and $b$, a number-like operator $N=ba$ can be defined, which is manifestly non self-adjoint, with $N^\dagger$ sharing with $N$ all its eigenvalues, $n=0,1,2,3,\ldots$. This is just the begin of the story for $\D$-PBs, which have a very rich structure.
 More on $\D$-PBs can be found in Section \ref{sectPBs}, while a rather complete (but not particularly recent) review is \cite{baginbagbook}. In \cite{bagrusso} it is shown that, under some mild extra condition, $a,b,N$ and their adjoint can be seen as elements of a $O^*$-algebra $\Lc^\dagger(\D)$ and, as such, we can multiply them, raise to (non-negative) integer powers, compute commutators, and so on.  This is useful for what follows.

\subsection{Extending CSusy}\label{sect2.2}

We are now ready to extend significantly the definition in (\ref{21}). In doing so, non self-adjoint operators will become relevant and natural. For that we start considering an Hilbert space $\Hil$, endowed with scalar product $\br.,.\kt$, and with an adjoint $\dagger$ connected to $\br.,.\kt$: $\br X^\dagger f,g\kt=\br f,Xg\kt$, $\forall f,g\in\Hil$. Let us further consider a suitable subspace $\D\subset\Hil$, and the $^*$-algebra  $\Lc^\dagger(\D)$ constructed as in \cite{aitbook}, see also Definition \ref{o*}. Here $\D$ can be constructed as in (\ref{add1}) for some operator $N_0$, or being a convenient dense subset of $\Hil$. Then

\begin{defn}\label{defecsusy}
	Let $d$, $c$, $r$ and $s$ be four elements of  $\Lc^\dagger(\D)$, and let $\gamma, \delta$ be two real numbers with $\delta>\gamma$. We say that $(d,c,r,s;\delta,\gamma)$ define an {\em extended coupled Susy} (ECSusy), if the following equalities are satisfied:
	\be
	\left\{
	\begin{array}{ll}
		dc=rs+\gamma\1,\\
		cd=sr+\delta\1,\\
	\end{array}
	\right.
	\label{25}\en
	
\end{defn}
Here, as usual, $\1$ is the identity operator on $\Hil$, and the formulas above could be understood as follows: $d(cf)=r(sf)+\gamma f$ and $c(df)=s(rf)+\delta f$, for all $f\in\D$. These equalities are both well defined since, if $f\in\D$, then $cf,df,sf,rf\in\D$ as well, and, therefore, f we also have $c(df)\in\D$, and so on.

In \cite{coupledsusy} it is shown that the two equalities in (\ref{21}) are really different: one can be satisfied while the other does not hold. This is what happens, for instance, if $a=\frac{1}{\sqrt{2}}\left(\frac{d}{dx}+x\right)$ and $b=\frac{1}{\sqrt{2}}\left(\frac{d}{dx}+x\right)e^{ix}$. Of course, it is not difficult to imagine that this difference between the two equations in (\ref{21}) is strengthen further in our case, i.e. for equations (\ref{25}). Stated differently, the two equations in (\ref{25}) are really different, and for this reason have to be considered together in the following.

Let us define the following operators, still in $\Lc^\dagger(\D)$:
\be
k_+=\frac{1}{\delta-\gamma}ds, \qquad k_-=\frac{1}{\delta-\gamma}rc, \qquad k_0=\frac{1}{\delta-\gamma}\left(dc-\frac{\gamma}{2}\,\1\right), 
\label{26}\en
and
\be
l_+=\frac{1}{\delta-\gamma}sd, \qquad l_-=\frac{1}{\delta-\gamma}cr, \qquad l_0=\frac{1}{\delta-\gamma}\left(sr+\frac{\delta}{2}\,\1\right).
\label{27}\en
Using (\ref{25}) it is easy to check that they obey the following commutation relations:
\be
[k_0,k_\pm]=\pm k_\pm, \qquad [k_+,k_-]=-2k_0,
\label{28}\en
as well as
\be
[l_0,l_\pm]=\pm l_\pm, \qquad [l_+,l_-]=-2l_0.
\label{29}\en
These look like the commutators in (\ref{23}), but with a big difference: $k_+$ and $l_+$ are not the adjoint of $k_-$ and $l_-$, and $k_0$ and $l_0$ are not self-adjoint. This gives us the possibility to introduce two other families of operators, $p_\alpha$ and $q_\alpha$, $\alpha=0,\pm$:
\be
p_0=k_0^\dagger, \qquad p_\pm=k_\mp^\dagger; \qquad\qquad q_0=l_0^\dagger, \qquad q_\pm=l_\mp^\dagger.
\label{210}\en
They satisfy the same commutators in (\ref{28}) and (\ref{29}):
\be
[p_0,p_\pm]=\pm p_\pm, \qquad [p_+,p_-]=-2p_0; \qquad [q_0,q_\pm]=\pm q_\pm, \qquad [q_+,q_-]=-2q_0.
\label{211}\en
Hence we conclude that (\ref{25}) implies the existence of four (in general) different triples of operators obeying the same commutators of an $\mathfrak{su}(1,1)$ Lie algebra, but with different relations under the adjoint operation. 

\subsection{Deformed $\mathfrak{su}(1,1)$ Lie algebra: a view to its eigenstates}\label{sectdla}

In the attempt to clarify the consequences of what deduced before, in this section we will deduce the main properties of three operators, $x_\pm$ and $x_0$, in $\Lc^\dagger(\D)$, satisfying  $[x_0,x_\pm]=\pm x_\pm$, and $[x_+,x_-]=-2x_0$, but with $x_+^\dagger\neq x_-$ and $x_0^\dagger\neq x_0$. In particular we will show that all the useful (for us)  results valid for $\mathfrak{su}(1,1)$ can be also deduced in the present situation.

First we put, with a slight abuse of notation, 
\be
x^2=x_0^2-\frac{1}{2}(x_+x_-+x_-x_+)=x_0^2+x_0-x_-x_+=x_0^2-x_0-x_+x_-.
\label{212}\en
We call it {\em an abuse} since $x^2$ is not really the square of an operator $x$ (to be identified) and, moreover, $x^2$ in (\ref{212}) is not even positive. We use this notation since it is somehow standard. In fact, the definition in (\ref{212}) is the one usually adopted in the literature for the {\em ordinary} $\mathfrak{su}(1,1)$ Lie algebra, and we borrow this terminology from that. The operator $x^2$ commutes with each $x_\alpha$: $[x^2,x_\alpha]=0$, for $\alpha=0,\pm$, and this is the main reason why it is so relevant for us.  Now, since in particular $x^2$ and $x_0$ commute, we can look for common eigenstates of these two operators. Using again the same notation adopted for ordinary $\mathfrak{su}(1,1)$, we assume the following: there exists a non zero vector $\Phi_{j,q_0}\in\D$ satisfying the following eigenvalue equations:
\be
\left\{
\begin{array}{ll}
	x^2\Phi_{j,q_0}=j(j+1)\Phi_{j,q_0},\\
	x_0\Phi_{j,q_0}=q_0\Phi_{j,q_0},\\
\end{array}
\right.
\label{213}\en
for some $j$ and $q_0$. We should stress that, in principle, there is no reason a priori to assume here that $j$ and $q_0$ are real or positive. This is because, as already observed, $x^2$ is not positive or self-adjoint, and $x_0$ is not self-adjoint, too. This makes in general more complicated to describe the range of values of $j$ and $q_0$. However, some useful result can still be found, as we will see. Using the commutation rules for $(x^2,x_\alpha)$ we deduce that
\be
\left\{
\begin{array}{ll}
	x^2(x_\pm\Phi_{j,q_0})=j(j+1)(x_\pm\Phi_{j,q_0}),\\
	x_0(x_\pm\Phi_{j,q_0})=(q_0\pm1)(x_\pm\Phi_{j,q_0}),\\
\end{array}
\right.
\label{214}\en
at least if $\Phi_{j,q_0}\notin \ker(x_\pm)$. This means that $x_\pm$ are ladder operators and, in particular, that $x_+$ is a raising while $x_-$ is a lowering operator. Using the same standard arguments for $\mathfrak{su}(1,1)$, we can also deduce that
\be
\left\{
\begin{array}{ll}
	x_+\Phi_{j,q_0}=(q_0-j)\Phi_{j,q_0+1},\\
	x_-\Phi_{j,q_0}=(q_0+j)\Phi_{j,q_0-1}.\\
\end{array}
\right.
\label{215}\en
These equations are in agreement with the fact that, as it is easy to check,
$$
[x_0,x_-x_+]=[x_0,x_+x_-]=0.
$$
In fact, from (\ref{215}) we see that $x_0$ and $x_-x_+$ have the same eigenvectors. The same is true for  $x_0$ and $x_+x_-$.

We have several possibilities: 

\vspace{2mm}

{\bf Case 1:--} for some $m\in\mathbb{N}_0=\mathbb{N}\cup\{0\}$ we have $x_-^{m-1}\Phi_{j,q_0}\neq0$ and $x_-^{m}\Phi_{j,q_0}=0$. In this case the set of eigenvalues of $x_0$, $\sigma(x_0)$, is bounded below: $\sigma(x_0)=\{q_0-m+1,q_0-m+2,q_0-m+3,\ldots\}$.

\vspace{2mm}

{\bf Case 2:--} for some $k\in\mathbb{N}_0=\mathbb{N}\cup\{0\}$ we have $x_+^{k-1}\Phi_{j,q_0}\neq0$ and $x_+^{k}\Phi_{j,q_0}=0$. In this case $\sigma(x_0)$, is bounded above: $\sigma(x_0)=\{\ldots,q_0+k-3,q_0+k-2,q_0+k-1\}$.

\vspace{2mm}

{\bf Case 3:--} both conditions above are true: for some $m\in\mathbb{N}_0=\mathbb{N}\cup\{0\}$ we have $x_-^{m-1}\Phi_{j,q_0}\neq0$ and $x_-^{m}\Phi_{j,q_0}=0$, and  for some $k\in\mathbb{N}_0=\mathbb{N}\cup\{0\}$ we have $x_+^{k-1}\Phi_{j,q_0}\neq0$ and $x_+^{k}\Phi_{j,q_0}=0$. In this case, of course, $\sigma(x_0)$, is bounded above and below: $\sigma(x_0)=\{q_0-m+1,q_0-m+2,\ldots,q_0+k-2,q_0+k-1\}$.

\vspace{2mm}

{\bf Case 4:--} neither Case 1, nor Case 2, hold. Then $\sigma(x_0)$ has no bound below and above.

\subsection{Back to ECSusy}\label{sectBECSusy}

We can use now the results of the previous section in the analysis of the operators introduced in Section \ref{sect2.2}. However, this will not be the only ingredient of the procedure we are going to propose. In fact, as we will see, the natural biorthonormality connected to the appearance of non self-adjoint number-like operators will play a relevant role. We first consider the operators $k_\alpha$, $\alpha=0,\pm$. As in (\ref{213}), we assume a non zero vector $\varphi_{j,q}\in\D$ exists, $j,q\in\mathbb{C}$, such that
\be
k^2\varphi_{j,q}=j(j+1)\varphi_{j,q}, \qquad k_0\varphi_{j,q}=q\varphi_{j,q}.
\label{216}\en
Here, as in (\ref{212}), $k^2=k_0^2+k_0-k_-k_+$, for instance. The operators $k_\pm$ act on $\varphi_{j,q}$ as ladder operators:
\be
k_+\varphi_{j,q}=(q-j)\varphi_{j,q+1}, \qquad k_-\varphi_{j,q}=(q+j)\varphi_{j,q-1}, 
\label{217}\en for all $\varphi_{j,q}\notin\ker(k_\pm)$. Let us now call $I_j$ the set of all the $q's$ for which $\varphi_{j,q}$ is not annihilated by at least one between $k_+$ and $k_-$: if $q\in I_j$, then $\varphi_{j,q}\notin\ker(k_+)$ or $\varphi_{j,q}\notin\ker(k_-)$, or both, and let $\F_\varphi(j):=\{\varphi_{j,q}, \,\forall q\in I_j\}$. Let then introduce $\E_j=l.s.\{\varphi_{j,q}, \, q\in I_j\}$, the linear span of the vectors in $\F_\varphi(j)$, and $\Hil_j$ the closure of $\E_j$, with respect to the norm of $\Hil$. Of course, $\Hil_j\subseteq\Hil$, for each fixed $j$. By construction, $\F_\varphi(j)$ is a basis for $\Hil_j$. Let $\F_\psi(j):=\{\psi_{j,q}, \,\forall q\in I_j\}$ be its unique biorthogonal basis, \cite{chri}. Then 
\be
\br\varphi_{j,q},\psi_{j,r}\kt=\delta_{q,r}, 
\label{218}\en
for all $q,r\in I_j$, and  $l.s.\{\psi_{j,q}, \, q\in I_j\}$ is dense in $\Hil_j$. Using (\ref{210}), it is possible to check the following eigenvalue and ladder equalities:
\be
\left\{
\begin{array}{ll}
	p_0\psi_{j,q}=\overline{q}\psi_{j,q},\\
	p_+\psi_{j,q}=\overline{(q+1+j)}\psi_{j,q+1},\\
	p_-\psi_{j,q}=\overline{(q-1-j)}\psi_{j,q+1},\\
\end{array}
\right.
\label{219}\en
at least if $\psi_{j,q}\notin\ker(p_\pm)$. 

\vspace{2mm}

{\bf Remark:--} These equations appear different from those deduced in (\ref{215}) and, in fact, this is so. In particular, the coefficients in the ladder equations are not those in (\ref{215}). The difference arises because the vectors $\{\psi_{j,q}\}$ are introduced here as the only family which is biorhonormal to the $\{\varphi_{j,q}\}$, which is more convenient, and more natural, for us. The other possibility would be to introduce, in analogy to what we have done in (\ref{216}), a family of eigenstate of $p^2$ and $p_0$, $\{\tilde\psi_{j,q}\}$, which, however, turns out to be biorthogonal, but not biorthonormal, to  $\{\varphi_{j,q}\}$. In other words, the difference we have with these different procedures is in the normalization of the states: $\psi_{j,q}$ and $\tilde{\psi}_{j,q}$ are proportional to each other. This is confirmed by the fact that, from (\ref{219}), we also deduce that $p^2\psi_{j,q}=j(j+1)\psi_{j,q}$. This aspect will appear clear in the analysis of $\D$-PBs, in Section \ref{sectPBs}, where these vectors will be explicitly computed and compared.

\vspace{2mm}

The biorthonormality between $\F_\varphi(j)$ and $\F_\psi(j)$ is connected to the fact that they are eigenstates of pairs of operators related by the adjoint map, as in (\ref{210}). The same formulas show that a similar relation exists between the operators $l_\alpha$ and $q_\alpha$. Also, comparing (\ref{26}) and (\ref{27}), we can see that, for instance, $l_\pm$ are a sort of supersymmetric version of $k_\pm$, meaning with this that the various operators are all factorized, and the order of the operators in, say, $l_\alpha$ is the opposite with respect to their order in $k_\alpha$. Hence the eigenstates of $(k_0,k^2)$ are also related to those of $(l_0,l^2)$, for example, as we will show next.

We begin by noticing that, as a consequence of (\ref{26}), (\ref{27}) and (\ref{210}), we can deduce the following intertwining relations:
\be
\left\{
\begin{array}{ll}
	sk_+=l_+s,\qquad\qquad k_+d=dl_+\\
	ck_-=l_-c,\qquad\qquad k_-r=rl_-\\
	r^\dagger p_+=q_+r^\dagger,\qquad\quad p_+c^\dagger=c^\dagger q_+\\
	d^\dagger p_-=q_-d^\dagger,\qquad\quad p_-s^\dagger=s^\dagger q_-,\\
\end{array}
\right.
\label{220}\en
as well as
\be
\left\{
\begin{array}{ll}
	l_0s=s\left(k_0+\frac{1}{2}\1\right)\\
	l_0c=c\left(k_0-\frac{1}{2}\1\right)\\
	rl_0=\left(k_0+\frac{1}{2}\1\right)r\\
	dl_0=\left(k_0-\frac{1}{2}\1\right)d,\\
\end{array}
\right.
\label{221}\en
and their adjoint which, recalling (\ref{210}), we can write as follows:
\be
\left\{
\begin{array}{ll}
	q_0r^\dagger=r^\dagger\left(p_0+\frac{1}{2}\1\right)\\
	q_0d^\dagger=d^\dagger\left(p_0-\frac{1}{2}\1\right)\\
	s^\dagger q_0=\left(p_0+\frac{1}{2}\1\right)s^\dagger\\
	c^\dagger q_0=\left(p_0-\frac{1}{2}\1\right)c^\dagger.\\
\end{array}
\right.
\label{222}\en

\vspace{2mm}

The equalities in (\ref{221}) and (\ref{222}) show that the eigenvalues of $l_0$ differ from those of $k_0$ by half integers, as those of $q_0$ from those of $p_0$. Indeed we have, considering a vector $\varphi_{j,q}$ with  $s\varphi_{j,q}\neq0$ and $c\varphi_{j,q}\neq0$,
$$
l_0\left(s\varphi_{j,q}\right)=s\left(k_0+\frac{1}{2}\1\right)\varphi_{j,q}=\left(q+\frac{1}{2}\right)\left(s\varphi_{j,q}\right).
$$
as well as
$$
l_0\left(c\varphi_{j,q}\right)=\left(q-\frac{1}{2}\right)\left(c\varphi_{j,q}\right).
$$
In other words, $s\varphi_{j,q}$ and $c\varphi_{j,q}$ are both eigenstates, with different eigenvalues, of $l_0$. Hence, as in (\ref{214}), they must be connected by the ladder operators $l_\pm$. However, we do not expect that the ladder equations are exactly those in (\ref{215}), since the eigenstates of $l_0$ are introduced using SUSY rather than considering the approach described in Section \ref{sectBECSusy}. Nonetheless, the vectors which are deduced here are proportional to those we would introduce using (\ref{213}), (\ref{214}) and (\ref{215}), $\chi_{j,q'}$, where $q'$ differs from $q$ for half-integers. The same equations (\ref{221}) allow to check that, if $r\chi_{j,q'}\neq0$ and $d\chi_{j,q'}\neq0$,
$$
k_0\left(r\chi_{j,q'}\right)=\left(q'-\frac{1}{2}\right)\left(r\chi_{j,q'}\right), \qquad k_0\left(d\chi_{j,q'}\right)=\left(q'+\frac{1}{2}\right)\left(d\chi_{j,q'}\right).
$$
This situation is clearly shown in Figure \ref{fig1} in a concrete situation which is deduced deforming $\D$-PBs, see Section \ref{sectDPBs}. More concretely, what we have just deduced here is {\em the right half part} of Figure \ref{fig1}. The left half part is connected with the consequences of the intertwining relations in (\ref{222}), consequences which are completely analogous to those discussed so far for $k_0$ and $l_0$. We will go back to Figure \ref{fig1} later on, when dealing with $\D$-PBs.

\section{A detailed example: $\D$-PBs}\label{sectPBs}

In this Section we will discuss in details how $\D$-PBs, see Definition \ref{def21}, fit into the present framework. In particular, we will show that a rather rich functional structure can be constructed, and generalized even more, as we will show in Section \ref{sectDPBs}.

We start this section recalling few facts related to Definition \ref{def21}, starting with the following working assumptions, valid in several physical systems considered along the years.

\vspace{2mm}

{\bf Assumption $\D$-pb 1.--}  there exists a non-zero $\varphi_{ 0}\in\D$ such that $a\,\varphi_{ 0}=0$.

\vspace{1mm}

{\bf Assumption $\D$-pb 2.--}  there exists a non-zero $\Psi_{ 0}\in\D$ such that $b^\dagger\,\Psi_{ 0}=0$.

\vspace{2mm}

The invariance of $\D$ under the action of $b$ and $a^\dagger$ implies 
that the vectors \be \varphi_n:=\frac{1}{\sqrt{n!}}\,b^n\varphi_0,\qquad \Psi_n:=\frac{1}{\sqrt{n!}}\,{a^\dagger}^n\Psi_0, \label{31}\en
$n\geq0$, are well defined and they all belong to $\D$ and, as a consequence, to the domain of $a^\sharp$, $b^\sharp$ and $N^\sharp$, where $N=ba$. Let us put $\F_\Psi=\{\Psi_{ n}, \,n\geq0\}$ and
$\F_\varphi=\{\varphi_{ n}, \,n\geq0\}$.
It is  simple to deduce the following lowering and raising relations:
\be
\left\{
\begin{array}{ll}
	b\,\varphi_n=\sqrt{n+1}\varphi_{n+1}, \qquad\qquad\quad\,\, n\geq 0,\\
	a\,\varphi_0=0,\quad a\varphi_n=\sqrt{n}\,\varphi_{n-1}, \qquad\,\, n\geq 1,\\
	a^\dagger\Psi_n=\sqrt{n+1}\Psi_{n+1}, \qquad\qquad\quad\, n\geq 0,\\
	b^\dagger\Psi_0=0,\quad b^\dagger\Psi_n=\sqrt{n}\,\Psi_{n-1}, \qquad n\geq 1,\\
\end{array}
\right.
\label{32}\en as well as the eigenvalue equations $N\varphi_n=n\varphi_n$ and  $N^\dagger\Psi_n=n\Psi_n$, $n\geq0$. Then,  if we choose the normalization of $\varphi_0$ and $\Psi_0$ in such a way $\left<\varphi_0,\Psi_0\right>=1$, we deduce that
\be \left<\varphi_n,\Psi_m\right>=\delta_{n,m}, \label{33}\en
for all $n, m\geq0$. Hence $\F_\Psi$ and $\F_\varphi$ are biorthonormal. In \cite{baginbagbook} it is shown that, in several quantum models, $\F_\Psi$ and $\F_\varphi$ are complete in $\Hil$, but they are not bases. However, they still produce useful resolution of the identity since they are always, at least in all the systems considered so far, $\G$-quasi bases, where $\G$ is some subspace dense in $\Hil$. This means that
for all $f$ and $g$ in $\G$,
\be
\left<f,g\right>=\sum_{n\geq0}\left<f,\varphi_n\right>\left<\Psi_n,g\right>=\sum_{n\geq0}\left<f,\Psi_n\right>\left<\varphi_n,g\right>.
\label{34}
\en

We refer to \cite{baginbagbook} for many more results and examples on $\D$-quasi bosons. Here, what is relevant for us, are the ladder properties described by (\ref{32}), and the fact that they produce a concrete, and highly non trivial, example of ECSusy.

In fact, let us fix the operators and the numbers $\gamma$ and $\delta$ in Definition \ref{defecsusy} as follows: $c=r=a$, $d=s=b$, $\delta=-\gamma=1$, where $a$ and $b$ satisfy Definition \ref{def21}. Hence the operators in (\ref{26}), (\ref{27}) and (\ref{210}) become
\be
k_+=l_+=\frac{1}{2}b^2, \qquad k_-=l_-=\frac{1}{2}a^2, \qquad k_0=l_0=\frac{1}{2}\left(N+\frac{1}{2}\1\right),
\label{35}\en
where $N=ba$, and
\be
p_+=q_+=\frac{1}{2}{a^\dagger}^2, \qquad p_-=q_-=\frac{1}{2}{b^\dagger}^2, \qquad p_0=q_0=\frac{1}{2}\left(N^\dagger+\frac{1}{2}\1\right).
\label{36}\en
It is clear that the four original families collapse to two. Formula (\ref{212}) produce further the following result:
\be
k^2=p^2=-\frac{3}{16}\1,
\label{37}\en
which, of course, commute with all the other operators, as expected. We notice that this formula clarifies what already observed in Section \ref{sectdla}, after formula (\ref{213}): despite of their "names", $k^2$ and $p^2$ are not positive operators. Formula (\ref{213}) is based on the assumption that a non zero eigenstate of $x^2$ and $x_0$ exists. In our situation,  such a vector can be easily found: in fact, if we consider the vacuum $\varphi_0$ introduced before, see Assumption $\D$-pb 1., we have $$k^2\varphi_0=-\frac{3}{16}\varphi_0, \qquad k_0\varphi_0=\frac{1}{4}\varphi_0.$$
Hence, comparing these with (\ref{213}), we have $q_0=\frac{1}{4}$ and $j(j+1)=-\frac{3}{16}$, that is $j=-\frac{1}{4}$ or $j=-\frac{3}{4}$. Because of formula (\ref{217}), and observing that $k_-\varphi_0=0$, we choose 
$j=-\frac{1}{4}$ and we define
\be
\varphi_{-\frac{1}{4},\frac{1}{4}}:=\varphi_0.
\label{38}\en
Hence we are in Case 1 of Section \ref{sectalgebras}, with $m=1$. In fact, since the spectrum of $N$ is the set $\mathbb{N}_0$, $\sigma(k_0)$ is bounded below.

If we act $m$ times with $k_+$ on $\varphi_{-\frac{1}{4},\frac{1}{4}}$, $m=1,2,3,\ldots$, formula (\ref{217}) produces
\be
\varphi_{-\frac{1}{4},m+\frac{1}{4}}=\frac{\sqrt{(2m)!}}{(2m-1)!!}\varphi_{2m},
\label{39}\en
where $\varphi_{2m}$ are those in (\ref{31}) and, with standard notation, $(2m-1)!!=1\cdot3\cdots(2m-3)\cdot(2m-1)$, with $0!!=(-1)!!=1$. Using (\ref{216}) and (\ref{217}), or with a direct check, we find
\be
k_0\varphi_{-\frac{1}{4},m+\frac{1}{4}}=\left(m+\frac{1}{4}\right)\varphi_{-\frac{1}{4},m+\frac{1}{4}},
\label{310}\en
and
\be
k_+\varphi_{-\frac{1}{4},m+\frac{1}{4}}=\left(m+\frac{1}{2}\right)\varphi_{-\frac{1}{4},m+\frac{5}{4}}, \qquad k_-\varphi_{-\frac{1}{4},m+\frac{1}{4}}=m\,\varphi_{-\frac{1}{4},m-\frac{3}{4}}.
\label{311}\en
In particular, this last equality is true only if $m\geq1$. If $m=0$ we have $k_-\varphi_{-\frac{1}{4},\frac{1}{4}}=k_-\varphi_0=0$, as already noticed.

According to Section \ref{sectBECSusy}, we can now define the set of linearly independent vectors $\F_\varphi^{(e)}\left(\frac{1}{4}\right)=\{\varphi_{-\frac{1}{4},m+\frac{1}{4}}, \, m=0,1,2,3,\ldots\}$, and the Hilbert space $\Hil_{-\frac{1}{4}}^{(e)}$, constructed by taking the closure of the linear span of its vectors. Here the suffix {\em e} stands for {\em even}, since only the vectors $\varphi_{2m}$ belong to $\F_\varphi^{(e)}\left(\frac{1}{4}\right)$. It is clear that $\Hil_{-\frac{1}{4}}^{(e)}\subset\Hil$, since all the vectors with odd index, $\varphi_{2m+1}$, which belong to $\Hil$, do not belong to $\Hil_{-\frac{1}{4}}^{(e)}$. Hence, the set 
$\F_\varphi^{(e)}\left(\frac{1}{4}\right)$ cannot be complete in $\Hil$, and, as a consequence, cannot be a basis for $\Hil$. Nevertheless, by construction, $\Hil_{-\frac{1}{4}}^{(e)}$ is an Hilbert space as well, and $\F_\varphi^{(e)}\left(\frac{1}{4}\right)$ is a basis for it. Then, see \cite{chri}, an unique biorthonormal basis $\F_\psi^{(e)}\left(\frac{1}{4}\right)=\{\psi_{-\frac{1}{4},m+\frac{1}{4}}, \, m=0,1,2,3,\ldots\}$ exists, such that
\be
\br\varphi_{-\frac{1}{4},m+\frac{1}{4}},\psi_{-\frac{1}{4},l+\frac{1}{4}}\kt=\delta_{m,l},
\label{312}\en
where the scalar product is the one in $\Hil$, and, for each $f\in \Hil_{-\frac{1}{4}}^{(e)}$,
\be
f=\sum_{m=0}^{\infty}\br\varphi_{-\frac{1}{4},m+\frac{1}{4}},f\kt\,\psi_{-\frac{1}{4},m+\frac{1}{4}}=\sum_{m=0}^{\infty}\br\psi_{-\frac{1}{4},m+\frac{1}{4}},f\kt\,\varphi_{-\frac{1}{4},m+\frac{1}{4}}.
\label{313}\en
From (\ref{39}) and (\ref{33}) it is clear that the vectors of this biorthonormal basis are the following:
\be
\psi_{-\frac{1}{4},m+\frac{1}{4}}=\frac{(2m-1)!!}{\sqrt{(2m)!}}\psi_{2m}.
\label{314}\en
Formulas (\ref{219}) can now be explicitly checked, and we get 
\be
p^2\psi_{-\frac{1}{4},m+\frac{1}{4}}=-\frac{3}{16}\psi_{-\frac{1}{4},m+\frac{1}{4}}, \qquad p_0\psi_{-\frac{1}{4},m+\frac{1}{4}}=\left(m+\frac{1}{4}\right)\psi_{-\frac{1}{4},m+\frac{1}{4}}, 
\label{315}\en
together with
\be
p_+\psi_{-\frac{1}{4},m+\frac{1}{4}}=\left(m+1\right)\psi_{-\frac{1}{4},m+\frac{5}{4}}, \qquad p_-\psi_{-\frac{1}{4},m+\frac{1}{4}}=\left(m-\frac{1}{2}\right)\psi_{-\frac{1}{4},m-\frac{3}{4}}.
\label{316}\en

Once more, we stress that the difference between these ladder equations and those in (\ref{311}) arises because, while the $\varphi_{-\frac{1}{4},m+\frac{1}{4}}$'s are introduced using directly the deformed $\mathfrak{su}(1,1)$ algebra, the $\psi_{-\frac{1}{4},m+\frac{1}{4}}$'s are just the unique basis which is biorthonormal to $\F_\varphi^{(e)}\left(\frac{1}{4}\right)$. However, as (\ref{315}) and (\ref{316}) show, these vectors are still eigenstates of $p^2$ and $p_0$, and obey interesting ladder equations with respect to $p_\pm$, which are slightly different from those in (\ref{215}).

Let us now consider the four intertwining relations (\ref{221}). In the case of $\D$-PBs, these correspond to the following two equalities:
\be
k_0b=b\left(k_0+\frac{1}{2}\1\right), \qquad k_0a=a\left(k_0-\frac{1}{2}\1\right).
\label{317}\en
The consequence of these equalities is widely considered in the literature: if $\rho$ is an eigenstate of $k_0$ with eigenvalue $E$, $k_0\rho=E\rho$, and if $a\rho$ and $b\rho$ are both non zero, then 
$$
k_0(a\rho)=\left(E-\frac{1}{2}\right)(a\rho), \qquad k_0(b\rho)=\left(E+\frac{1}{2}\right)(b\rho),
$$
which means that $a\rho$ and $b\rho$ are both eigenstates of $k_0$, but with two shifted (and different) eigenvalues, $E\pm\frac{1}{2}$. Now, since (\ref{310}) shows that the eigenvalues related to different vectors $\varphi_{-\frac{1}{4},m+\frac{1}{4}}$ and $\varphi_{-\frac{1}{4},l+\frac{1}{4}}$ differ for integer quantities, we conclude that neither $a\varphi_{-\frac{1}{4},m+\frac{1}{4}}$, nor $b\varphi_{-\frac{1}{4},m+\frac{1}{4}}$, can still be of the same form $\varphi_{-\frac{1}{4},l+\frac{1}{4}}$, for any $l\in\mathbb{N}_0$. And, in fact, this can be explicitly checked, since
\be
a\varphi_{-\frac{1}{4},m+\frac{1}{4}}=\sqrt{2m}\frac{\sqrt{(2m)!}}{(2m-1)!!}\,\varphi_{2m-1}, \qquad b\varphi_{-\frac{1}{4},m+\frac{1}{4}}=\frac{\sqrt{(2m+1)!}}{(2m-1)!!}\,\varphi_{2m+1}, 
\label{318}\en
with the agreement that $\varphi_{-1}=0$. Let us now define
\be
\varphi_{-\frac{1}{4},m+\frac{3}{4}}:=b\varphi_{-\frac{1}{4},m+\frac{1}{4}}=\frac{\sqrt{(2m+1)!}}{(2m-1)!!}\,\varphi_{2m+1},
\label{319}\en
for all $m\geq0$. The reason for calling this vector in this way is because $\varphi_{-\frac{1}{4},m+\frac{3}{4}}$ is an eigenstate of $k_0$ with eigenvalue $m+\frac{3}{4}$, as expected because of our previous analysis on $b\rho$:
\be
k_0\varphi_{-\frac{1}{4},m+\frac{3}{4}}=\left(m+\frac{3}{4}\right)\varphi_{-\frac{1}{4},m+\frac{3}{4}},
\label{320}\en
We further deduce the following raising and lowering relations:
\be
k_+\varphi_{-\frac{1}{4},m+\frac{3}{4}}=\left(m+\frac{1}{2}\right)\varphi_{-\frac{1}{4},m+\frac{7}{4}}, \qquad k_-\varphi_{-\frac{1}{4},m+\frac{3}{4}}=m\,\frac{2m+1}{2m-1}\,\varphi_{-\frac{1}{4},m-\frac{1}{4}},
\label{321}\en
with the agreement that $\varphi_{-\frac{1}{4},-\frac{1}{4}}=0$.

\vspace{2mm}

{\bf Remark:--} It is now possible to check explicitly that (\ref{321}) are different from (\ref{215}). The reason is that, as it is clear, we are extending here the strategy described in Section \ref{sectBECSusy}, where a single family of operators were considered. Now, in fact, we are considering 4 families of operators satisfying the same commutator rules, and we are putting all these operators together, keeping biorthonormality as our main requirement, since we think this is {\em the most natural way}, mainly for technical reasons.  

For completeness, let us briefly consider what happens if we repeat for the operators $p_\alpha$ what we have done for the $k_\alpha$. In analogy with (\ref{38}), we put $\tilde\psi_{-\frac{1}{4},\frac{1}{4}}=\psi_0$, since $b^\dagger\psi_0=0$. Then, using the first equation in (\ref{215}) (rather than the second in (\ref{219})), $p_+\tilde\psi_{j,q}=(q-j)\tilde\psi_{j,q+1}$, we deduce that
$$
\tilde\psi_{-\frac{1}{4},m+\frac{1}{4}}=\frac{(2m-1)!!}{\sqrt{(2m)!}}\,\psi_{2m}=\frac{(2m)!}{((2m-1)!!)^2}\psi_{-\frac{1}{4},m+\frac{1}{4}},
$$
which shows the difference in the normalizations between the $\tilde\psi_{-\frac{1}{4},m+\frac{1}{4}}$ and the $\psi_{-\frac{1}{4},m+\frac{1}{4}}$. Now, while $\tilde\psi_{-\frac{1}{4},m+\frac{1}{4}}$ satisfies the analogous of formulas (\ref{215}), $\psi_{-\frac{1}{4},m+\frac{1}{4}}$ satisfies (\ref{316}), which is slightly different. On the other hand, while this last vector satisfies (\ref{312}), $\tilde\psi_{-\frac{1}{4},m+\frac{1}{4}}$ does not. In conclusion, we prefer to keep biorthonormality of the sets we work with, rather than using (\ref{215}) several times. But this is, of course, just a matter of personal taste.

\vspace{3mm}

It is clear that, in the same way in which $a$ and $b$ map $\varphi_{-\frac{1}{4},m+\frac{1}{4}}$ into some $\varphi_{-\frac{1}{4},l+\frac{3}{4}}$, they also map these last vectors into the first ones. More explicitly, we have
\be
a\varphi_{-\frac{1}{4},m+\frac{3}{4}}=\sqrt{2m+1}\,\frac{\sqrt{(2m+1)!}}{(2m-1)!!}\,\varphi_{2m}, \qquad b\varphi_{-\frac{1}{4},m+\frac{3}{4}}=\frac{\sqrt{(2m+2)!}}{(2m-1)!!}\,\varphi_{2m+2}, 
\label{322}\en
for all $m\geq0$. Notice that the vectors in the RHS of these equalities are proportional to $\varphi_{-\frac{1}{4},m+\frac{1}{4}}$ and to $\varphi_{-\frac{1}{4},m+\frac{5}{4}}$, see (\ref{29}).

In analogy with what we have done before, we introduce now the set $\F_\varphi^{(o)}\left(\frac{1}{4}\right)=\{\varphi_{-\frac{1}{4},m+\frac{3}{4}}, \, m=0,1,2,3,\ldots\}$, where {\em o} stands for {\em odd}, and the Hilbert space $\Hil_{-\frac{1}{4}}^{(o)}$, constructed by taking the closure of the linear span of its vectors. It is clear that $\Hil_{-\frac{1}{4}}^{(e)}\cap\Hil_{-\frac{1}{4}}^{(o)}=\emptyset$, and that, together, $\F_\varphi\left(\frac{1}{4}\right):=\F_\varphi^{(e)}\left(\frac{1}{4}\right)\cup\F_\varphi^{(o)}\left(\frac{1}{4}\right)$ is complete in $\Hil$, at least if the set $\F_\varphi$ is complete, which is always the case in all the concrete examples in the literature, so far. In particular, if the $\D$-PBs are {\em regular}, see \cite{baginbagbook}, $\F_\varphi$ and $\F_\psi$ are biorthonormal Riesz bases. Hence $\F_\varphi\left(\frac{1}{4}\right)$ is a Riesz basis as well.

Now, since $\F_\varphi^{(o)}\left(\frac{1}{4}\right)$ is a basis for $\Hil_{-\frac{1}{4}}^{(o)}$, we can introduce an unique biorthonormal  basis $\F_\psi^{(o)}\left(\frac{1}{4}\right)=\{\psi_{-\frac{1}{4},m+\frac{3}{4}}, \, m=0,1,2,3,\ldots\}$, whose vectors can be easily identified using (\ref{319}) and (\ref{33}). We have
\be
\psi_{-\frac{1}{4},m+\frac{3}{4}}=\frac{(2m-1)!!}{\sqrt{(2m+1)!}}\,\psi_{2m+1}=\frac{1}{2m+1}a^\dagger \psi_{-\frac{1}{4},m+\frac{1}{4}}.
\label{323}\en
It may be interesting to notice the difference in the normalization between $\psi_{-\frac{1}{4},m+\frac{3}{4}}$ and $\varphi_{-\frac{1}{4},m+\frac{3}{4}}$, in terms of their $m+\frac{1}{4}$ counterparts, see (\ref{319}) and (\ref{323}). This difference arises because we want to maintain biorthonormality of the vectors. In fact, with the choice in (\ref{323}) we get
\be
\br\varphi_{-\frac{1}{4},m+\frac{3}{4}},\psi_{-\frac{1}{4},l+\frac{3}{4}}\kt=\delta_{m,l},
\label{324}\en
where the scalar product is the one in $\Hil$, and, for each $f\in \Hil_{-\frac{1}{4}}^{(o)}$,
\be
f=\sum_{m=0}^{\infty}\br\varphi_{-\frac{1}{4},m+\frac{3}{4}},f\kt\,\psi_{-\frac{1}{4},m+\frac{3}{4}}=\sum_{m=0}^{\infty}\br\psi_{-\frac{1}{4},m+\frac{3}{4}},f\kt\,\varphi_{-\frac{1}{4},m+\frac{3}{4}}.
\label{325}\en
Repeating then what we have done for $\Hil^{(e)}$, we can define the set $\F_\psi^{(o)}\left(\frac{1}{4}\right)=\{\psi_{-\frac{1}{4},m+\frac{3}{4}}, \, m=0,1,2,3,\ldots\}$, and observe that
$\F_\psi\left(\frac{1}{4}\right):=\F_\psi^{(e)}\left(\frac{1}{4}\right)\cup\F_\psi^{(o)}\left(\frac{1}{4}\right)$ is complete in $\Hil$, or it is even a Riesz basis for $\Hil$, depending on the nature of the $\D$-PBs we are considering. More in detail, if we now introduce the families $\F_\Phi=\{\Phi_k, \, k\geq0\}$ and $\F_\xi=\{\xi_k, \, k\geq0\}$, where
$$
\Phi_k=\left\{
\begin{array}{ll}
\varphi_{-\frac{1}{4},j+\frac{1}{4}}, \qquad\quad\,\, \mbox{if } k=2j,\\
\varphi_{-\frac{1}{4},j+\frac{3}{4}}, \qquad\quad\,\, \mbox{if } k=2j+1,\\
\end{array}
\right.\quad\mbox{ and }\quad \xi_k=\left\{
\begin{array}{ll}
\psi_{-\frac{1}{4},j+\frac{1}{4}}, \qquad\quad\,\, \mbox{if } k=2j,\\
\psi_{-\frac{1}{4},j+\frac{3}{4}}, \qquad\quad\,\, \mbox{if } k=2j+1,\\
\end{array}
\right.
$$
$k\geq0$, we can check that $\br\Phi_k,\xi_l\kt=\delta_{k,l}$, and that, $\forall f,g\in\D$,
$$
\sum_{k=0}^{\infty}\br f,\Phi_k\kt\br\xi_k,g\kt= \sum_{k=0}^{\infty}\br f,\varphi_k\kt\br\psi_k,g\kt, \qquad \sum_{k=0}^{\infty}\br f,\xi_k\kt\br\Phi_k,g\kt= \sum_{k=0}^{\infty}\br f,\psi_k\kt\br\varphi_k,g\kt.
$$
These equalities imply that $\F_\Phi$ and $\F_\xi$ are biorthonormal, and, \cite{baginbagbook}, that they are $\D$-quasi bases
if and only if $\F_\varphi$ and $\F_\psi$ are $\D$-quasi bases. This property, useful to deduce several mathematical properties of the system, as already stated, is always true in all the physical systems where $\D$-PBs have been shown to appear so far, \cite{baginbagbook,bagthmp}.

We end this section with Tables \ref{table1} and \ref{table2} which contain several useful formulas involving all the vectors introduced in this section.

\begin{table}[h]
	
	\begin{tabular}{||l|l||}
		\hline\hline
		\rule[-4mm]{0mm}{1.2cm}
		$\varphi_{-\frac{1}{4},m+\frac{1}{4}}=\frac{\sqrt{(2m)!}}{(2m-1)!!}\varphi_{2m}$ & $\varphi_{-\frac{1}{4},m+\frac{3}{4}}=b\varphi_{-\frac{1}{4},m+\frac{1}{4}}=\frac{\sqrt{(2m+1)!}}{(2m-1)!!}\varphi_{2m+1}$\\
		\hline
		\rule[-4mm]{0mm}{1.0cm}
		$\psi_{-\frac{1}{4},m+\frac{1}{4}}=\frac{(2m-1)!!}{\sqrt{(2m)!}}\psi_{2m}$ & $\psi_{-\frac{1}{4},m+\frac{3}{4}}=\frac{a^\dagger}{2m+1}\psi_{-\frac{1}{4},m+\frac{1}{4}}=\frac{(2m-1)!!}{\sqrt{(2m+1)!}}\psi_{2m+1}$\\
		\hline\hline
		\rule[-4mm]{0mm}{1.2cm}
		$a\varphi_{-\frac{1}{4},m+\frac{1}{4}}=\sqrt{2m}\frac{\sqrt{(2m)!}}{(2m-1)!!}\varphi_{2m-1}=\frac{2m}{2m-1}\varphi_{-\frac{1}{4},m-\frac{1}{4}}$ & $a\varphi_{-\frac{1}{4},m+\frac{3}{4}}=\sqrt{2m+1}\frac{\sqrt{(2m+1)!}}{(2m-1)!!}\varphi_{2m}=(2m+1)\varphi_{-\frac{1}{4},m+\frac{1}{4}}$  \\
		\hline
		\rule[-4mm]{0mm}{1.2cm}
		$b\varphi_{-\frac{1}{4},m+\frac{1}{4}}=\frac{\sqrt{(2m+1)!}}{(2m-1)!!}\varphi_{2m+1}=\varphi_{-\frac{1}{4},m+\frac{3}{4}}$ & $b\varphi_{-\frac{1}{4},m+\frac{3}{4}}=\frac{\sqrt{(2m+2)!}}{(2m-1)!!}\varphi_{2m+2}=(2m+1)\varphi_{-\frac{1}{4},m+\frac{5}{4}}$  \\
		\hline
		\rule[-4mm]{0mm}{1.2cm}
		$a^\dagger\psi_{-\frac{1}{4},m+\frac{1}{4}}=\frac{\sqrt{(2m+1)!}}{(2m+1)!!}\psi_{2m+1}=(2m+1)\psi_{-\frac{1}{4},m+\frac{3}{4}}$ & $a^\dagger\psi_{-\frac{1}{4},m+\frac{3}{4}}=\sqrt{2m+2}\frac{(2m-1)!!}{\sqrt{(2m+1)!}}\psi_{2m+2}=\frac{2m+2}{2m+1}\varphi_{-\frac{1}{4},m+\frac{5}{4}}$  \\
		\hline
		\rule[-4mm]{0mm}{1.2cm}
		$b^\dagger\psi_{-\frac{1}{4},m+\frac{1}{4}}=\frac{(2m-1)!!}{\sqrt{(2m-1)!}}\psi_{2m+1}=(2m-1)\psi_{-\frac{1}{4},m-\frac{1}{4}}$ & $b^\dagger\psi_{-\frac{1}{4},m+\frac{3}{4}}=\frac{(2m-1)!!}{\sqrt{(2m)!}}\psi_{2m}=\psi_{-\frac{1}{4},m+\frac{1}{4}}$  \\
		\hline\hline
		
	\end{tabular}

	\caption{Useful formulas for $\D$-PBs, part 1}\label{table1}
\end{table}

\begin{table}[h]

\begin{tabular}{||l|l|l||}
	\hline\hline
	\rule[-4mm]{0mm}{1.2cm}
	$k_0\varphi_{-\frac{1}{4},m+\frac{1}{4}}=\left(m+\frac{1}{4}\right)\varphi_{-\frac{1}{4},m+\frac{1}{4}}$&$k_+\varphi_{-\frac{1}{4},m+\frac{1}{4}}=\left(m+\frac{1}{2}\right)\varphi_{-\frac{1}{4},m+\frac{5}{4}}$&$k_-\varphi_{-\frac{1}{4},m+\frac{1}{4}}=m\,\varphi_{-\frac{1}{4},m-\frac{3}{4}}$\\
	\hline
	\rule[-4mm]{0mm}{1.2cm}
		$p_0\psi_{-\frac{1}{4},m+\frac{1}{4}}=\left(m+\frac{1}{4}\right)\psi_{-\frac{1}{4},m+\frac{1}{4}}$&$p_+\psi_{-\frac{1}{4},m+\frac{1}{4}}=\left(m+1\right)\psi_{-\frac{1}{4},m+\frac{5}{4}}$&$p_-\psi_{-\frac{1}{4},m+\frac{1}{4}}=\left(m-\frac{1}{2}\right)\,\psi_{-\frac{1}{4},m-\frac{3}{4}}$\\
	\hline
	\rule[-4mm]{0mm}{1.2cm}
	$k_0\varphi_{-\frac{1}{4},m+\frac{3}{4}}=\left(m+\frac{3}{4}\right)\varphi_{-\frac{1}{4},m+\frac{3}{4}}$&$k_+\varphi_{-\frac{1}{4},m+\frac{3}{4}}=\left(m+\frac{1}{2}\right)\varphi_{-\frac{1}{4},m+\frac{7}{4}}$&$k_-\varphi_{-\frac{1}{4},m+\frac{3}{4}}=m\,\frac{2m+1}{2m-1}\,\varphi_{-\frac{1}{4},m-\frac{1}{4}}$\\
	\hline
	\rule[-4mm]{0mm}{1.2cm}
		$p_0\psi_{-\frac{1}{4},m+\frac{3}{4}}=\left(m+\frac{3}{4}\right)\psi_{-\frac{1}{4},m+\frac{3}{4}}$&$p_+\psi_{-\frac{1}{4},m+\frac{3}{4}}=\left(m+1\right)\frac{2m+3}{2m+1}\psi_{-\frac{1}{4},m+\frac{7}{4}}$&$p_-\psi_{-\frac{1}{4},m+\frac{3}{4}}=\left(m-\frac{1}{2}\right)\,\psi_{-\frac{1}{4},m-\frac{1}{4}}$\\
	\hline\hline

\end{tabular}

	\caption{Useful formulas for $\D$-PBs, part 2}\label{table2}

\end{table}

\subsection{Deformed $\D$-PBs}\label{sectDPBs}

In Section \ref{sectPBs} we were somehow forced to simplify the structure of ECSusy because the four original families of operators collapse into two. In this section we will  sketch how to deform $\D$-PBs to find four different operators satisfying Definition \ref{defecsusy}. It will be obvious that this can be done in very different ways, each of which produces a concrete example of ECSusy.

Let $a$ and $b$ be $\D$-pseudo-bosonic in the sense of Definition \ref{def21}. We work, as always, under the conditions which ensure that they belong, together with their adjoints, to $\Lc^\dagger(\D)$, for some suitable $\D$. Let now $S, T\in\Lc^\dagger(\D)$ be two invertible operators, with  $S^{-1}, T^{-1}\in\Lc^\dagger(\D)$. In the following we will assume that ${T^{-1}}^\dagger={T^{\dagger}}^{-1}$ and ${S^{-1}}^\dagger={S^{\dagger}}^{-1}$. Conditions for these to be satisfied can be found, for instance, in \cite{hiroshi}, Lemma 3.2. If we define now
\be
c=SaT^{-1}, \qquad s=SbT^{-1}, \qquad d=TbS^{-1}, \qquad r=TaS^{-1},
\label{326}\en 
it is clear that these operators, which are all in $\Lc^\dagger(\D)$, satisfy (\ref{25}) with $\delta=-\gamma=1$. Using (\ref{26}), (\ref{27}) and (\ref{210}), together with (\ref{326}), we find that
\be
\tilde k_\alpha=Tk_\alpha T^{-1}, \qquad \tilde l_\alpha=Sk_\alpha S^{-1}, \qquad \tilde p_\alpha={T^{-1}}^\dagger p_\alpha T^{\dagger}, \qquad   \tilde q_\alpha={S^{-1}}^\dagger p_\alpha S^{\dagger},
\label{327}\en
where $\alpha=0,\pm$ and where the {\em un-tilted} operators $k_\alpha$ and $p_\alpha$ are those in (\ref{35}) and (\ref{36}). Recalling now that $\varphi_{-\frac{1}{4},m+\frac{1}{4}}, \psi_{-\frac{1}{4},m+\frac{1}{4}}, \varphi_{-\frac{1}{4},m+\frac{3}{4}}, \psi_{-\frac{1}{4},m+\frac{3}{4}}\in\D$, for all $m=0,1,2,3,\ldots$, it follows that the following vectors are in $\D$ as well:
\be
\tilde\varphi_{-\frac{1}{4},m+\frac{1}{4}}=T\varphi_{-\frac{1}{4},m+\frac{1}{4}}; \qquad \tilde\psi_{-\frac{1}{4},m+\frac{1}{4}}={T^{-1}}^\dagger\psi_{-\frac{1}{4},m+\frac{1}{4}}; 
\label{328}\en
and 
\be 
\tilde\chi_{-\frac{1}{4},m+\frac{3}{4}}=S\varphi_{-\frac{1}{4},m+\frac{3}{4}}; \qquad
\tilde\eta_{-\frac{1}{4},m+\frac{3}{4}}={S^{-1}}^\dagger\psi_{-\frac{1}{4},m+\frac{3}{4}}.
\label{329}\en
They are eigenstates respectively of $\tilde k_0$ and $\tilde p_0$, with eigenvalue $m+\frac{1}{4}$, and of $\tilde l_0$ and $\tilde q_0$, with eigenvalue $m+\frac{3}{4}$. Moreover, they satisfy the following ladder equations:
\be
\left\{
\begin{array}{ll}
	\tilde k_+\tilde\varphi_{-\frac{1}{4},m+\frac{1}{4}}=\left(m+\frac{1}{2}\right)\tilde\varphi_{-\frac{1}{4},m+\frac{5}{4}}, \hspace{2.4cm} \tilde k_-\tilde\varphi_{-\frac{1}{4},m+\frac{1}{4}}=m\tilde\varphi_{-\frac{1}{4},m-\frac{3}{4}},\\
	\tilde p_+\tilde\psi_{-\frac{1}{4},m+\frac{1}{4}}=\left(m+1\right)\tilde\psi_{-\frac{1}{4},m+\frac{5}{4}}, \hspace{2.4cm} \tilde p_-\tilde\psi_{-\frac{1}{4},m+\frac{1}{4}}=\left(m-\frac{1}{2}\right)\tilde\psi_{-\frac{1}{4},m-\frac{3}{4}},\\
	\tilde l_+\tilde\chi_{-\frac{1}{4},m+\frac{3}{4}}=\left(m+\frac{1}{2}\right)\tilde\chi_{-\frac{1}{4},m+\frac{7}{4}}, \hspace{2.4cm} \tilde l_-\tilde\chi_{-\frac{1}{4},m+\frac{3}{4}}=m\,\frac{2m+1}{2m-1}\tilde\chi_{-\frac{1}{4},m-\frac{1}{4}},\\
	\tilde q_+\tilde\eta_{-\frac{1}{4},m+\frac{3}{4}}=\left(m+1\right)\,\frac{2m+3}{2m+1}\tilde\eta_{-\frac{1}{4},m+\frac{7}{4}}, \hspace{1.6cm} \tilde q_-\tilde\eta_{-\frac{1}{4},m+\frac{3}{4}}=\left(m-\frac{1}{2}\right)\tilde\eta_{-\frac{1}{4},m-\frac{1}{4}},\\
\end{array}
\right.
\label{330}\en
for every $m$ for which the lowering operators do not destroy the state. Also, they are biorthonormal in pairs, meaning that
\be
\br\tilde\varphi_{-\frac{1}{4},m+\frac{1}{4}},\tilde\psi_{-\frac{1}{4},l+\frac{1}{4}}\kt=\br\tilde\chi_{-\frac{1}{4},m+\frac{3}{4}},\tilde\eta_{-\frac{1}{4},l+\frac{3}{4}}\kt=\delta_{m,l},
\label{331}\en
for all $m,l\in\mathbb{N}_0$, while, if $S$ and $T$ are not chosen in some special way,  we get, for instance, $\br\tilde\varphi_{-\frac{1}{4},m+\frac{1}{4}},\tilde\eta_{-\frac{1}{4},l+\frac{3}{4}}\kt\neq0$. More than this: if we repeat here the same construction leading to the definition of the families $\F_\Phi$ and $\F_\xi$, see Section \ref{sectDPBs}, we do not get $\D$-quasi bases for $\Hil$, except for special choices of $S$ and $T$. A trivial possibility is when $S$ and $T$ are different, but proportional. Another more interesting case, can be set up if the various eigenvectors have definite parity. In this case, biorthormality of the sets is ensured in both $S$ and $T$ are multiplication operators: $Sf(x)=s(x)f(x)$ and $Tf(x)=t(x)f(x)$, $f(x)\in\D$, if $s(x)$ and $t(x)$ are even functions.

\begin{figure}
	\centering
	\includegraphics[width=\columnwidth]{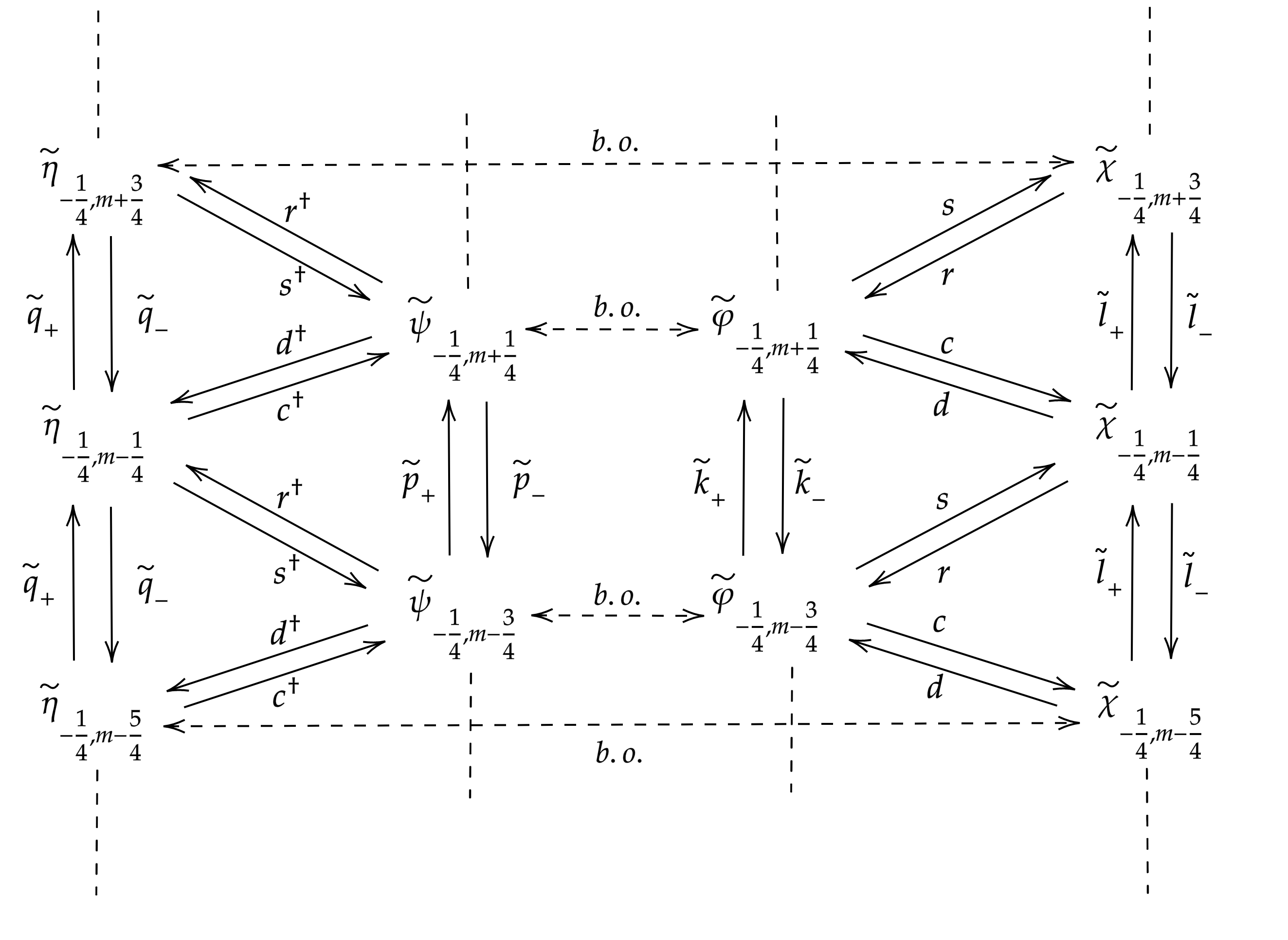}
	\caption{The role of the ladder operators, of the adjoint map, and of the biorthonormality of the eigenvectors.} 
	\label{fig1}
\end{figure}

 The whole situation is summarized in Figure \ref{fig1}: the vertical lines describe the action of the various ladder operators for the different families of vectors. The dashed horizontal lines connect the various biorthonormal families (and, in fact, this is the meaning of {\em b.o.} over the lines). The various slanted lines represent the extension of the results in Table \ref{table1} in the present case. For instance, we can easily check that
\be
s\,\tilde\varphi_{-\frac{1}{4},m+\frac{1}{4}}=\tilde\chi_{-\frac{1}{4},m+\frac{3}{4}}, \qquad c\,\tilde\varphi_{-\frac{1}{4},m+\frac{1}{4}}=\frac{2m}{2m-1}\tilde\chi_{-\frac{1}{4},m-\frac{1}{4}},
\label{add2}\en
and so on. Incidentally we observe that the slanted lines are only meant to show which vector is mapped into which other vector, but does not give any information on the related coefficients: for instance in the two cases just mentioned in (\ref{add2}), in the first case the coefficient is just $1$, while in the second is $\frac{2m}{2m-1}$. This difference is not made explicit in the Figure. It is also useful to stress that Figure \ref{fig1} does not really refer only to the present, very particular, example of ECSusy, deduced as a deformation of $\D$-PBs, but it is absolutely general. Also, it is not hard to understand what Figure \ref{fig1} becomes in the case described in Section \ref{sectPBs}, since, for instance, $c$ and $r$ collapse (and coincide with $a$), and so on.

\subsubsection{A concrete choice of $a$, $b$, $S$ and $T$}

The example we discuss here is connected to a {\em shifted harmonic oscillator}. The relevant aspect of the example is that the shift is complex, and this produces a manifestly non self-adjoint Hamiltonian. We refer to \cite{petr4,bag2013pra} for several results on this and similar systems. 

Let $c_0$ be the standard bosonic annihilation operators on $\Lc^2(\mathbb{R})$, $c_0=\frac{1}{\sqrt{2}}\left(x+\frac{d}{dx}\right)$, $[c_0,c_0^\dagger]=\1$. Since $c_0$ and $c_0^\dagger$ are in $\Lc^\dagger(\Sc(\mathbb{R}))$, where $\Sc(\mathbb{R})$ is the Schwartz space of the test functions, this commutator is well defined in $\Lc^\dagger(\Sc(\mathbb{R}))$. Also, if we call $N_0=c_0^\dagger c_0$, then $H_0=N_0+\frac{1}{2}\1$ is the (self-adjoint) Hamiltonian of the quantum harmonic oscillator. It is well known how its eigenvectors can be constructed: if $e_0(x)\in\Sc(\mathbb{R})$ satisfies $c_0\,e_0(x)=0$, then 
$$
e_n(x)=\frac{1}{\sqrt{n!}}\,(c_0^\dagger)^ne_0(x)=\frac{1}{\sqrt{2^nn!}\,\pi^{1/4}}\,H_n(x)e^{-x^2/2},
$$  
where $H_n(x)$ is the $n-$th Hermite polynomial, and we have $H_0e_n(x)=E_ne_n(x)$, with $E_n=n+\frac{1}{2}$, $n=0,1,2,3\ldots$.

Now, let us introduce, for a fixed $\alpha\in\mathbb{R}$, an operator $V_\alpha$ in the following way:
$$
(V_\alpha f)(x)=f(x-i\alpha),
$$
for all $f(x)\in D(V_\alpha)=\{h(x)\in\Lc^2(\mathbb{R}): \, h(x-i\alpha)\in\Lc^2(\mathbb{R})\}$. This set is dense in $\Lc^2(\mathbb{R})$, since it contains $\D$, the linear span of the $e_n(x)$'s, whose set $\F_e=\{e_n(x)\}$ is an orthonormal basis for $\Lc^2(\mathbb{R})$. Notice that $\D\subset\Sc(\mathbb{R})$. The inclusion $\D\subseteq D(V_\alpha)$ follows from the fact that
$$
|e_n(x-i\alpha)|^2=\frac{e^{\alpha^2}}{2^nn!\sqrt{\pi}}\,H_n(x+i\alpha)H_n(x-i\alpha)e^{-x^2},
$$
which is integrable since $H_n(x+i\alpha)H_n(x-i\alpha)$ is a (real and non negative) polynomial of degree $2n$. It is clear that $V_\alpha$ is invertible, and that, for all $f(x)\in\D$, $(V_\alpha^{-1}f)(x)=f(x+i\alpha)$. It is also easy to check that, $\forall f(x), g(x)\in\D$, $\br V_\alpha\,f,g\kt=\br f,V_\alpha g\kt$ and $\br V_\alpha^{-1}\,f,g\kt=\br f,V_\alpha^{-1} g\kt$. Also, $\br V_\alpha\,f,V_\alpha^{-1}g\kt=\br f, g\kt$. These equalities allow us to check that the two families
$$
\F_\varphi=\{\varphi_n(x)=(V_\alpha e_n)(x)=e_n(x-i\alpha)\}, \qquad \F_\psi=\{\psi_n(x)=(V_\alpha^{-1} e_n)(x)=e_n(x+i\alpha)\},
$$
are biorthonormal and $\D$ quasi-bases:
$$
\br f,g\kt=\sum_{n=0}^{\infty}\br f,\varphi_n\kt \br \psi_n,g\kt=\sum_{n=0}^{\infty}\br f,\psi_n\kt \br \varphi_n,g\kt,
$$
for all $f,g\in\D$. To introduce the $\D$-pseudo-bosonic ladder operators for $\F_\varphi$ we first observe that $V_\alpha c_0 V_\alpha^{-1}$ and $V_\alpha c_0^\dagger V_\alpha^{-1}$ are well defined on $\D$. In fact, taking $f(x)=\sum_{l=0}^{N}k_le_l(x)\in\D$, it follows first of all that $(V_\alpha^{-1}f)(x)=\sum_{l=0}^{N}\tilde k_lH_l(x+i\alpha)e^{-(x+i\alpha)^2/2}$, where $\tilde k_k=\frac{k_l}{2^ll!\sqrt{\pi}}$. Now, since each polynomial of degree $M$ with complex coefficients, $P_M(x)$, can be written as a linear combination of the first $M$ Hermite polynomials, $P_M(x)=\sum_{j=0}^Mc_jH_j(x)$ for some complex $\{c_j\}$, and since $\sum_{l=0}^N\tilde k_l xH_l(x+i\alpha)$ and  $\sum_{l=0}^N\tilde k_l (H_l'(x+i\alpha)-(x+i\alpha)H_l(x+i\alpha))$ are both polynomials of degree $N+1$, we deduce that
$$
c_0 V_\alpha^{-1}f(x)=\sum_{l=0}^{N+1}k_l^+H_l(x)e^{-(x+i\alpha)^2/2}, \qquad c_0^\dagger V_\alpha^{-1}f(x)=\sum_{l=0}^{N+1}k_l^-H_l(x)e^{-(x+i\alpha)^2/2},
$$
for some properly chosen set of coefficients $k_l^\pm$. It is clear that these two sums do not belong to $\D$, but they both belong to $D(V_\alpha)$, and we get, for instance
$$
V_\alpha c_0 V_\alpha^{-1}f(x)=\sum_{l=0}^{N+1}k_l^+H_l(x-i\alpha)e^{-x^2/2}=\sum_{l=0}^{N+1}q_l^+H_l(x)e^{-x^2/2},
$$
for some $q_l^+$ to be identified\footnote{It is clear that this (and other) identification is not relevant for us, here. We are only interested in checking that the operators are well defined, not really in computing their action. This is done in a different way.}. This suggests to introduce $a=V_\alpha c_0 V_\alpha^{-1}$, which is densely defined, since $D(a)\supseteq \D$. Similarly, we  introduce also $b=V_\alpha c_0^\dagger V_\alpha^{-1}$, and we find that $D(b)\supseteq\D$. 

To the same conclusion we can arrive
 using the Baker-Campbell-Hausdorff formula. In fact, the operator $V_\alpha$ coincides on $\D$ with $e^{\alpha p}$, $p$ being the momentum operator, and therefore
\be
a=V_\alpha c_0 V_{\alpha}^{-1}=c_0-\frac{i\alpha}{\sqrt{2}}, \qquad b=V_\alpha c_0^\dagger V_{\alpha}^{-1}=c_0^\dagger-\frac{i\alpha}{\sqrt{2}}.
\label{332}\en
which are in agreement with the fact that $D(a), D(b)\supseteq\D$, and with the fact that, as we have seen before, $a,b:\D\rightarrow\D$.

It is clear that $b^\dagger\neq a$, for $\alpha\neq0$, and that $[a,b]=\1$. The adjoints of $a$ and $b$ are the raising and the lowering operators for $\F_\psi$, \cite{baginbagbook}.

Now, the easiest example we can construct is the one we obtain by choosing $S=V_\sigma$ and $T=V_\tau$, $\sigma,\tau>0$ fixed, in (\ref{326}). They are both invertible, and we find, for instance,
$$
c=V_{\sigma+\alpha}c_0V_{-(\tau+\alpha)}, \quad d=V_{\tau+\alpha}c_0^\dagger V_{-(\sigma+\alpha)}, \quad s=V_{\sigma+\alpha}c_0^\dagger V_{-(\tau+\alpha)}, \quad r=V_{\tau+\alpha}c_0 V_{-(\tau+\alpha)}.
$$
Using (\ref{332}) we can rewrite, say, $c$, as follows: 
$$
c=V_{\sigma-\tau}\left(c_0-\frac{i(\alpha+\tau)}{\sqrt{2}}\right).
$$ 
Hence, taking $f(x)\in\D$ and observing that $(c_0-\frac{i(\alpha+\tau)}{\sqrt{2}})f(x)\in\D$ as well, we conclude that $cf(x)$ is well defined. Similar formulas can be established for $d$, $s$ and $r$.

We can also rewrite formulas (\ref{327}), (\ref{328}) and (\ref{329}). For instance, $\tilde k_+=V_{\alpha+\tau}\left(\frac{1}{2}{c_0^\dagger}^2\right)V_{-(\alpha+\tau)}$, while the {\em tilted} vectors look as follows:
$$
\tilde\varphi_{-\frac{1}{4},m+\frac{1}{4}}=\frac{\sqrt{(2m)!}}{(2m-1)!!}\,e_{2m}(x-i(\alpha+\tau)); \quad \tilde\psi_{-\frac{1}{4},m+\frac{1}{4}}=\frac{(2m-1)!!}{\sqrt{(2m)!}}\,e_{2m}(x+i(\alpha+\tau)); 
$$
and 
$$
\tilde\chi_{-\frac{1}{4},m+\frac{3}{4}}=\frac{\sqrt{(2m+1)!}}{(2m-1)!!}\,e_{2m+1}(x-i(\alpha+\sigma)); \quad \tilde\eta_{-\frac{1}{4},m+\frac{3}{4}}=\frac{(2m-1)!!}{\sqrt{(2m+1)!}}\,e_{2m+1}(x+i(\alpha+\sigma)).
$$

It may be worth remarking that, even in presence of biorthogonal sets, the basis property of these sets is not granted, in general, due to the fact that it does not even hold for ordinary pseudo-bosons. In fact, what we have always found in concrete examples considered along the years is that the different families of functions connected by our ladder operators are indeed complete in $\Hil$, and $\G$-quasi bases (for properly chosen $\G$), but not necessarily bases.

\section{Conclusions}\label{sectconcl}

In this paper we have introduced and analyzed the notion of ECSusy, deforming and extending some original ideas proposed in \cite{coupledsusy}. We have shown that, in doing this extension, how a deformed $\mathfrak{su}(1,1)$ Lie algebra can be introduced. We have also shown that this algebra, where, among other differences with the {\em standard} one, the ladder operators are not connected by the adjoint map, most of the usual results on $\mathfrak{su}(1,1)$ can be established, at least those related to the eigenstates of the number-like operator and to the ladder equations. $\D$-PBs have been considered in this perspective, and we have shown how to construct concrete examples of ECSusy. In particular, the role of biorthonormal sets, ladder operators and SUSY has been analyzed in detail.

\section*{Acknowledgements}

The author acknowledges partial support from Palermo University and from G.N.F.M. of the INdAM.

%
%
%

%
%

\end{document}